\documentclass[sigconf]{acmart}

\AtBeginDocument{%
  \providecommand\BibTeX{{%
    \normalfont B\kern-0.5em{\scshape i\kern-0.25em b}\kern-0.8em\TeX}}}

\copyrightyear{2024}
\acmYear{2024}
\setcopyright{acmlicensed}\acmConference[PEARC '24]{Practice and Experience in Advanced Research Computing}{July 21--25, 2024}{Providence, RI, USA}
\acmBooktitle{Practice and Experience in Advanced Research Computing (PEARC '24), July 21--25, 2024, Providence, RI, USA}
\acmDOI{10.1145/3626203.3670561}
\acmISBN{979-8-4007-0419-2/24/07}

\usepackage{multirow}
\usepackage{multicol}
\usepackage{xcolor}
\usepackage{colortbl}
\usepackage{listings} 
\usepackage{natbib}
\usepackage{arydshln} 

\begin{document}


\title[Automatic BLAS Offloading on Unified Memory Architecture]{Automatic BLAS Offloading on Unified Memory Architecture: A Study on NVIDIA Grace-Hopper}
\author{Junjie Li}
\email{jli@tacc.utexas.edu}
\orcid{0000-0002-1051-5927}
\affiliation{%
  \institution{Texas Advanced Computing Center, \\The University of Texas at Austin}
  \streetaddress{10100 Burnet Rd}
  \city{Austin}
  \state{Texas}
  \country{USA}
  \postcode{78758}
}

\author{Yinzhi Wang}
\email{iwang@tacc.utexas.edu}
\orcid{0000-0001-8505-0223}
\affiliation{%
  \institution{Texas Advanced Computing Center, \\The University of Texas at Austin}
  \streetaddress{10100 Burnet Rd}
  \city{Austin}
  \state{Texas}
  \country{USA}
  \postcode{78758}
}

\author{Xiao Liang}
\email{liangstein@psc.edu}
\orcid{0000-0002-7882-3571}
\affiliation{%
  \institution{Pittsburgh Supercomputing Center, \\Carnegie Mellon University}
  \streetaddress{300 S. Craig St.}
  \city{Pittsburgh}
  \state{Pennsylvania}
  \country{USA}
  \postcode{15213}
}

\author{Hang Liu}
\email{hliu@tacc.utexas.edu}
\orcid{0000-0002-3486-7863}
\affiliation{%
  \institution{Texas Advanced Computing Center, \\The University of Texas at Austin}
  \streetaddress{10100 Burnet Rd}
  \city{Austin}
  \state{Texas}
  \country{USA}
  \postcode{78758}
}

\renewcommand{\shortauthors}{J. Li and Y. Wang, et al.}

\begin{abstract}

Porting codes to GPU often requires major efforts. 
While several tools exist for automatically offload numerical libraries such as BLAS and LAPACK, they often prove impractical due to the high cost of mandatory data transfer.
The new unified memory architecture 
in NVIDIA Grace-Hopper allows high bandwidth cache-coherent memory access of all memory from both CPU and GPU, potentially eliminating bottleneck faced in conventional architecture.  
This breakthrough opens up new avenues for application development and porting strategies. 
In this study, we introduce a new tool for automatic BLAS offload, the tool leverages the high speed cache coherent NVLink C2C interconnect in  Grace-Hopper, and enables performant GPU offload for BLAS heavy applications with no code changes or recompilation. 
The tool was tested on two quantum chemistry or physics codes, great performance benefits were observed. 

\end{abstract}




\maketitle

\section{Introduction}
\label{sec:intro}

As computational demands escalate, the necessity of harnessing the computing power of Graphics Processing Units (GPUs) becomes increasingly pressing. 
GPUs offer unparalleled potential for accelerating computations in various scientific and computational domains. 
However, porting CPU codes to GPU isn't technically trivial, especially for large codes, legacy codes and codes with complex workflow.  
Additionally, it's been challenging for many academic researchers to get funded for such porting efforts.   
Considering many scientific applications utilize abstracted numerical libraries,  
there has been many attempts to automate these library calls for offload which are outlined in \ref{sec:previous-blas-offload}, 
but they offer limited performance benefit due to the complex traditional CPU-GPU memory architecture, 
and therefore not well adopted for practical use. 

The emergence of Unified Memory Architecture (UMA) in recent GPU designs, such as NVIDIA Grace-Hopper superchip and AMD M300A APU, opens up new possibilities for software design. 
With UMA providing a cache coherent unified memory access for both CPU and GPU, innovative data management strategies can potentially yield performant automatic offload tools.


In this paper, we focus on the influential linear algebra library BLAS, 
introducing a novel tool for automatic offload of CPU BLAS calls 
to the GPU without requiring user code modifications or recompilation. 
The tool is publicly available at \cite{scilib-accel} . 
Leveraging the UMA architecture, 
our tool offers several data management strategies for optimal performance. 
Section \ref{sec:review} provides a review of the UMA architecture 
and related BLAS offload work, while Section \ref{sec:implementation} elaborates on the implementation details of our tool. 
We then apply the tool to two BLAS-heavy scientific computing codes, 
presenting and discussing the results in Section \ref{sec:result}. 
Finally, we conclude in Section \ref{sec:conclusion}.

\section{Background and Related Work}
\label{sec:review}
 
\subsection{Coherent memory in NVIDIA Grace-Hopper} 
\label{sec:uma}

In conventional architectures, GPU and host memory exist in separate spaces, preventing direct access between CPU and GPU memory. 
To alleviate this hurdle, CUDA managed memory introduced since CUDA 6 allows sharing memory pointers, but implicit page migration is performed by the CUDA runtime behind the scene. 
In contrast, the Grace-Hopper superchip features closely integrated CPU and GPU units along with LPDDR5x and HBM3e memory subsystems, connected by the high-bandwidth and cache-coherent NVLink Chip-2-Chip (C2C) interconnect\cite{gh-whitepaper}. 
This technology enables a unified memory space where both CPU and GPU can access memory without page movement. From the OS perspective, these memories resemble two NUMA domains, akin to memories in a two-socket CPU system.

\subsection{Previous automatic BLAS offload attempts}
\label{sec:previous-blas-offload}


Numerous attempts have been made to automatically accelerate CPU BLAS calls since the early adoption of GPUs in HPC. 
 Cray LIBSCI\_ACC\cite{cray-libsci-acc, cray-libsci-acc-slide}, available for over a decade, was deployed on the Titan supercomputer, supporting selected BLAS, LAPACK, and ScaLAPACK routines for offload when the module is loaded. 
 Similarly, IBM ESSL\cite{ibm-essl-offload} is capable of automatically offloading selected BLAS, LAPACK, and FFTW calls once libesslsmpcuda is linked. 
 NVIDIA's NVBLAS\cite{nvidia-nvblas} serves as a drop-in replacement for CPU BLAS calls, allowing users to configure host BLAS libraries and selected routines for offload. By $LD\_PRELOAD$ NVBLAS, dynamically linked CPU BLAS is replaced without relinking. 
 These tools make offload decisions at runtime based on workload sizes and handles data movement automatically. 
 Overall, these libraries are tailored for conventional GPU architectures, and frequent data movement is unavoidable, therefore suffers poor performance for small and medium sized matrix math in real workloads.
 
\section{Implementation}
\label{sec:implementation}

Our tool is designed for the ease of use. 
Similar to NVBLAS, the tool is provided as a shared library (.so) file, and user only need to $LD\_PRELOAD$ the .so file. 
Several environmental variables can be set to choose which BLAS routine to be offloaded, offload criteria for matrix size, level of debug outputs, and data management strategy.  
The functionality is achieved by intercepting BLAS symbols.   
As level-3 BLAS calls are most compute intensive which best suites GPU, only level-3 BLAS functions are implemented so far. 

\subsection{Symbol Interception} 
Symbol interception is achieved via a trampoline-based Dynamic Binary Instrumentation (DBI) approach: 
a small piece of assembly code is inserted into the original function, enabling it to jump to a trampoline function. 
This trampoline function preserves the overwritten bytes by the extra jump instruction and executes customized code before returning to the original program. 

In the current context, we intercept BLAS calls where the trampoline function maintains the same signature as the original function.
This results in minimal overhead, as the register data for function arguments remains undisturbed. 
This mechanism finds extensive use in profilers, and here we use the PEAK\cite{peak} lightweight profiler framework we have developed,  ensuring portability across various architectures, including but not limited to x86 and ARM.
Also note that our DBI approach applies to both dynamically and statically linked BLAS, 
while other tools like NVBLAS, which works by resolving runtime library dependency, only work for dynamically linked BLAS.

\subsection{Data Movement}
Managing data movement is often the most critical part of GPU porting as data transfer speed has been a limiting factor.  
Here we provide three data management strategies taking advantage of the Grace-Hopper architecture. 
Note that in Grace-Hopper, LPPDDR5 and HBM are in NUMA 0 and NUMA 1 respectively, and both can be accessed by all processors in a cache-coherent fashion.  
\vspace{-1mm}
\begin{itemize}
    \item {Strategy 1, memory copies}: 
     This is the most intuitive strategy and  used by other tools. 
    Upon entry of a BLAS call,   input matrices are copied from host  memory to GPU memory,  
    then resultant matrix is copied back after cuBLAS call. 
    This strategy works on all GPUs including the conventional PCIe-based cards, 
    but at the cost of frequent data movement.    
    
    \item {Strategy 2, unified memory access}: 
    Since LPDDR5 and HBM are physically unified with cache-coherent NVLink C2C, 
    CPU malloced matrix pointers can be passed directly to cuBLAS calls.  
    One can use numactl to let all memory resident on HBM, 
    or recompile both the application and this tool with -gpu=unified flag in NVHPC compiler to allow CUDA runtime to automatic decide data residency.

    \item {\bf Strategy 3, automatic data migration}:
    Recognizing that CPU access to HBM is slower than to LPDDR5 (see section \ref{sec:result-dgemm}), allocating all memory on HBM as in Strategy 2 can slow down CPU code execution.
    Here, we designed a new {\bf first-touch type of data migration} scheme in which matrices are moved to HBM (by migrating pages) upon their first use by cuBLAS and they remain resident on the GPU until deallocation. 
\end{itemize}
\vspace{-1mm}
{\bf Strategy 3} overcomes all issues in Strategy 1 and 2. While the best strategy could be application dependent but {Strategy 3} should be optimal in most cases.

\section{Performance Testings and Discussion}  
\label{sec:result}

We focus on exploring the performance on the novel Grace-Hopper (also known as GH200) unified memory architecture. 
Our Grace-Hopper superchip contains one 72-core Grace CPU with 480GB LPDDR5x host memory and one H100 GPU with 96GB HBM3e memory.  
Note that the 480GB model has lower bandwidth than models with less LPDDR5x, so performance is expected to be better on those models.  
The GPU driver on GH200 is the latest 550.54.14 and CUDA version is 12.4. 
Power can be dynamically allocated within the superchip, and TDP of the entire chip is 1000W.  
The latest NVHPC 24.3 compiler and math libraries in the compiler suite are used throughout the testing.  

Certain tests were also done on the regular PCIe-based H100 cards for comparison.   
This test system is equipped with dual-socket AMD EPYC 7454 (Milan) CPU and two H100-PCIe GPUs each with 80GB of HBM3,
but we are only using one CPU and one GPU for fair comparison with a single Grace-Hopper superhcip. 
GPU driver version 535.104.12 and CUDA version is 12.2.  The TDP for H100-PCIe is 350W. 
As the CPU is x86, we used intel compiler plus MKL on the CPU, and used cuBLAS from NVHPC 24.3 suite on the GPU. 

In all the following test cases, matrix multiplication with problem size $(mnk)^{1/3}>500$ will be offloaded which is proven to be appropriate to benefit from GPU. 
The optimal offload threshold depends on the relative compute capability of GPU versus CPU and latency, and the value can be adjusted by user. 

\subsection{Matrix Multiplication (dgemm)}  
\label{sec:result-dgemm} 

\begin{table}[ht]
\centering
\centering
\caption{STREAM Bandwidth on GH200 (GB/s)} 
\vspace{-3mm}
\label{tab:stream}
\begin{tabular}{l|c|c c}
\hline
               & & LPDDR5 & HBM \\
\hline
\multirow{4}{*}{CPU{\scriptsize$^\star$} }  
               & Copy  &312.71 &   129.61 \\
               & Mul   &305.65 &   130.62 \\
               & Add   &314.47 &   125.93 \\
               & Triad &314.59 &   125.94 \\
\cline{1-4}
\multirow{4}{*}{GPU}
               & Copy  &318.26 &   3421.95 \\
               & Scale &318.37 &   3417.83 \\
               & Add   &477.91 &   3741.64 \\
               & Triad &477.24 &   3739.18 \\
\hline
\end{tabular}
\end{table}
\begin{table}[ht]
\centering\captionsetup{justification=centering}
 \caption{DGEMM Runtime with Unified Memory \\(transA='T', transB='N', M=32, N=2400, K=93536)}\vspace{-3mm}
\label{tab:gemm-uma}
\begin{tabular}{l|c c}
\hline
& CPU (72C) & GPU \\
& (dgemm) & (cublasDgemm) \\
\hline
LPDDR5 &  19.7 ms &  19.7 ms \\
HBM &   24.9 ms & 0.84 ms \\
\hline
\end{tabular}

\end{table}

\begin{table}[ht]
\centering
\captionsetup{justification=centering} 
\caption{ Runtime of cublasDgemm w/ cudaMemCpy (transA='T', transB='N', M=32, N=2400, K=93536)}\vspace{-3mm}
\label{tab:cublas-w-cpy}
\resizebox{\columnwidth}{!}{%
\begin{tabular}{l|llcc}
\hline
                  & GH200 & H100-PCIe & GH200 & H100-PCIe \\
                  \hline
Offload Strategy              & Strategy 1   & Strategy 1     & NVBLAS  & NVBLAS\\
\hline
Total time                    &   5.50ms     &   32.80ms     & 54.8 ms & 134.0ms \\
\quad 1. cudaMemcpy$^\dagger$ & \quad\ 4.96 ms & \quad 31.79ms  & -       & - \\
\quad 2. cublasDgemm          & \quad\ 0.52 ms & \quad\ 0.99ms   & -       & -\\
\quad 3. other                & \quad\ 0.02 ms & \quad\ 0.02ms  & -       & -\\
\hline
\multicolumn{5}{l}{\scriptsize $^\dagger$ Including copying matrices A, B and C to GPU memory and C back to host memory. } \\
\end{tabular}%
}
\end{table}


We tested dgemm call for a matrix size (M=32, N=2400, K=93536) which often appears in the PARSEC application tested in the next subsection. 
Table \ref{tab:gemm-uma} shows performance of using unified memory access with zero data copy (Strategy 2). 
It's seen that GPU on HBM is tremendously faster than CPU on LPDDR5, and GPU on LPDDR5 performs similarly to CPU on LPDDR5. 
However, run CPU on HBM is much slower. This is understandable by the STREAM bandwidth in Table \ref{tab:stream}, the bandwidth of CPU accessing HBM is quite limited, even slower than accessing LPDDR5.  
Table \ref{tab:cublas-w-cpy} shows performance of using explicit cudaMemcpy for every cuBLAS call (Strategy 1), 
while GH200 and H100-PCIe are both extremely fast in performing dgemm, data transfer in GH200 is much faster than transferring through the PCIe lanes.  
Note that the cublasDgemm time on cudamalloced memory  (0.52ms) in Table \ref{tab:cublas-w-cpy} is consistently faster than running on host allocated HBM (0.84ms) in Table \ref{tab:gemm-uma}. 
NVBLAS is also tested on GH200, but significantly slower than our tool. 
The breakdown timing was measured by internal timer in our tool, so no detailed timing for NVBLAS is available. 
It is also worth mentioning that CUDA pinned memory is enabled in both our tool and NVBLAS.


\subsection{Application Test 1: PARSEC} 
\label{sec:parsec}
PARSEC (Pseudopotential Algorithm for Real-Space Electronic Calculations)\cite{parsec1,parsec2} is a package designed to perform Density Functional Theory (DFT) calculations of solids and molecules.
It solves the Kohn–Sham equations in real space, without the use of explicit basis sets.
 Our test case calculates energy of a Silicon nanocrystal $Si_{1947}H_{604}$, boundary sphere radius is set to 50 bohr, grid spacing is 0.9 bohr, the calculation is limited to two self-consistent field steps to reduce benchmark cost, but performance characteristics of a fully converged calculation is identical. 

 PARSEC has been a pure CPU code since it was first developed, it heavily relies on ScaLAPACK, and the dgemm calls made from ScaLAPACK can easily exceed 50\% of runtime in real use cases. 
 With the help of our tool, this is the first time PARSEC performantly runs on a GPU.

\definecolor{intelblue}{RGB}{0, 199, 253} 
\definecolor{nvidiagreen}{RGB}{118, 185, 0}
\begin{table*}[ht]
\caption{Performance of Different Offload Strategies for PARSEC}
\vspace{-3mm}
\label{tab:parsec-offload-time}
\begin{tabular}{l|c|c|c|l}
\hline
Hardware & Optimal setup & Wall time  & dgemm+data$^\dagger$  & BLAS Strategy \\
       &  (mpi x omp)  & (seconds) &   (seconds)   &  \\
\hline
EPYC 9454 (48C)   & 16x3  & 574.9  & 315.8  & MKL  \\
EPYC 9454 + H100-PCIe & 16x3 & 915.6 & - & NVBLAS \\
EPYC 9454 + H100-PCIe & 16x3 & 511.9 & 269.8  & Strategy 1 \\  
\hline
GH200: CPU (72C)   & 72x1 & 824.6 & 562.0 & OpenBLAS in NVHPC \\
\hdashline
\multirow{5}{*}
{GH200: GPU}   & 16x4 &  508.0 & 310.8 &Strategy 1 \\  
               & 16x4 & 290.1 & 23.9                                                  &Strategy2, HBM pinned \\
               & 16x4 & 682.0 & 414.5 &Strategy 2, -gpu=unified \\
               & 16x4 & {\bf 246.6} & 36.7 & Strategy 3 \\ 
               &  -    & error  & - &   NVBLAS \\
\hline
\multicolumn{5}{l}{\scriptsize $^\dagger$ max value over all MPI ranks } \\
\end{tabular}
\end{table*}

This code is comprehensively tested on two platforms using both our tool and NVBLAS for offload, results are shown in Table \ref{tab:parsec-offload-time}.
On H100-PCIe card where only Strategy 1 can be used,  our tool performs slightly faster than the pure CPU run 
while NVBLAS is nearly 2x slower. 
Data movement consumes more than half of the runtime and prevents automatic offload from being competitive. 
On GH200, Strategy 2 combined with HBM residence through numactl offers significant speedup over the Grace CPU run, and also much faster than a single AMD Geona CPU. 
As discussed in Section \ref{sec:result-dgemm}, CPU accessing HBM is slower than accessing LPDDR5, therefore pinning all memory onto HBM has some performance penalty.  
This negative impact is resolved by Strategy 3 where matrices are moved only once upon first use by GPU. 
Strategy 3 further reduces the runtime and delivers 3.3x speedup comparing to Grace CPU.  
NVBLAS is tested as well, but fails in the middle of execution with no helpful error message.  

Strategy 1 on GH200 and H100-PCIe performs similarly due to differences in ScaLAPACK implementations. 
We used MKL on x86, and libscalapack in NVHPC on GH200.
Function pdgemm in MKL makes much fewer number of dgemm calls resulting 14.6 TB total data movement at 64GB/s PCIe bandwidth, while the implementations in NVHPC made 101 TB data movement at bandwidth of 370GB/s.  
Performance on GH200 could improve with optimized ScaLAPACK.

 We further investigated reason of Strategy 3 being so great here, each matrix migrated by the tool is reused 445 times in subsequent dgemm calls, and total time spent on page migration is about 10s while total dgemm time is reduced from nearly 600s on CPU to only about 26s on GPU. 

With the optimal automatically offload strategy, dgemm accounts for less than 15\% of the total runtime , compared to 68\% on Grace or 54\% on EPYC 9454. 
Using GPU eliminates the dgemm bottleneck, and additional optimization on the CPU code is needed to achieve better GPU utilization.

\subsection{Application Test 2: MuST}

\begin{table*}[!ht] 

\caption{Performance of Different Offload Strategies for MuST}\vspace{-3mm}
\label{tab:must-offload-time}
\begin{tabular}{l|c|c|c|l}
\hline
Hardware & Optimal setup & Wall time & zgemm+data$^\dagger$  & BLAS Strategy \\
         &  (mpi x omp)  & (seconds) & (seconds)  & \\
\hline
GH200: CPU &  56x1  &  127.5   &  83.4 & NVPL \\ 
\hdashline
\multirow{5}{*}{GH200: GPU} &  28x2   &  {\bf\ 57.5} & - & native GPU implementation \\
                            &  28x2   &  {80.8} & 34.0 & Strategy 1 \\
                            &  28x2   &  \ 74.5      & 14.4  &Strategy2, pinned on HBM \\
                            & 28x2   &  \ {\bf 62.8}      & 18.3  &Strategy 3 \\
                            &  -   &  error          & - &NVBLAS \\

\hline
\multicolumn{5}{l}{\scriptsize $^\dagger$ max value over all MPI ranks } \\
\end{tabular}
\end{table*}

MuST (Multiple Scattering Theory)\cite{must1,must2} is a package designed to perform electronic structure calculations, it solves the Kohn-Sham equation by solving the Green's function.
In contrast to solving the wave-function, it can perform KKR-CPA calculations for random structures and LSMS calculations for large systems with linear scalability to the system size.
A large portion of solving the Green's function is calculating the inverse of the KKR matrix. 
The code has a heavy dependency on zgemm operations which often exceeds 60\% of runtime on CPU. 

MuST natively supports GPU offload, 
the native method implemented in MuST offloads the matrix inverse onto a GPU by calling cuSOLVER.
Our test case calculates a CoCrFeMnNi supercell alloy using the LSMS method. Total atom number in the supercell is 56 and  concentration of each element is identical. The calculation is limited to 3 self-consistent steps to reduce benchmark cost.

This test case heavily relies on zgemm routine as shown in Table \ref{tab:must-offload-time}. 
Thorough testing was conducted on GH200 with various offload strategies. 
Again Strategy 3 is the most effective for automatic offload and performs nearly as fast as the native GPU implementation using cuSOLVER.  There is also a very high matrix re-use rate that makes matrix migration time very small. 
Both of these offload implementations are more than twice as fast as running on the pure Grace CPU.
NVBLAS was attempted, but crashes again with no meaningful error message.  

In regard of the zgemm dependence, it is 83s or 65\% of total runtime on CPU, while drastically reduced to 18.3s or 29\% in the best auto offload setup and is no longer a major hot spot.

\section{Conclusion} 
\label{sec:conclusion}
In this paper, we report a new tool that can intercept Level 3 BLAS symbols in a CPU code, 
and automatically perform GPU offload using GPU-enabled BLAS along with several choices of data management strategies, especially a newly proposed first-touch type of matrix migration scheme. 
The tool is fine-tuned to achieve minimum overhead and take full advantage of the new unified memory architecture with cache coherent NVLink C2C. 
Performance tests on BLAS heavy codes show significant speedup comparing to CPU code, and the tool far outperforms NVBLAS auto-offload tool provided by NVIDIA  on not only Grace-Hopper  but also on the conventional PCIe-based GPU. 
The tool is useful for users and code developers to quickly explore the potential benefits of using GPU
and have a quick start on the new architecture.  


\begin{acks}
This work is supported by the National Science Foundation through awards OAC-2402542, OAC-1854828, and OAC-2139536. 
\end{acks}

\bibliographystyle{ACM-Reference-Format}
\bibliography{reference}



\end{document}